\documentclass[rnote]{aa}
\bibpunct{(}{)}{;}{a}{}{,} 
\usepackage{graphicx}
\begin{document}
\title{On the relation between the non-flat cosmological models and the elliptic integral of first kind}
\titlerunning{On the relation between the ...}  

\author{A. M\'esz\'aros \inst{1}
\and
J. \v{R}\'{\i}pa \inst{2}
}

\offprints{A. M\'esz\'aros}

\institute{Charles University in Prague, Faculty of Mathematics and Physics, Astronomical Institute,\\
V Hole\v{s}ovi\v{c}k\'ach 2, CZ 180 00 Prague 8, Czech Republic\\
\email{meszaros@cesnet.cz}\\
\and
Institute of Basic Science, Natural Sciences Campus, Sungkyunkwan University,\\
Engineering Building 2, 2066 Seobu-ro, Jangan-gu, Suwon, Gyeonggi-do, 440-746, Korea\\
\email{ripa.jakub@gmail.com}
}

\date{Received October 22, 2014; accepted October 25, 2014}

\abstract {Recently we have found an analytic integration formula that describes the dependence 
of the luminosity distance on the redshift for the flat cosmological model with a non-zero
cosmological constant.}
{The purpose of this article is to search for a similar relation for the non-flat models.}
{A Taylor series was used.}
{The elliptic integral of the first kind can indeed be applied.}
{The result shows that a combination of the numerical integration with the previous analytic formula can also be useful
for the nearly flat cosmological models.}

\maketitle

\section{Introduction}

In our recent paper \citep{meri13}, we have shown that 
the standard cosmological dependence of the luminosity distance $d_L(z)$ on the redshift $z$ 
can also be described analytically for a non-zero cosmological constant, especially if the
Universe is spatially flat. In this result the elliptic integral of the first kind plays a key role.

The question emerges immediately of whether the restriction to the flat cosmological model 
is needed or does the elliptic integral 
of the first kind play a role also for non-flat models with non-zero cosmological constant? 
This note studies these questions.

\section{A recapitulation of the known relations}
 
We start with \citep{car92}
$$ d_{PM}(z) = \frac{c}{H_0 \sqrt{|\Omega_k|}} \times \;\;\;\;\;\;\;\;$$
\begin{equation}
  {\rm sinn}
 \left\{\sqrt{|\Omega_k|} \int_{0}^{z}
                   \frac{dz'}{\sqrt{(1+z')^3 \Omega_M  + (1+z')^2\Omega_k +\Omega_{\Lambda}}}\right\}.
\end{equation}
In this equation, $c$ is the speed of light in vacuum, $H_0$ is the Hubble-constant, and
$\Omega_k + \Omega_M + \Omega_{\Lambda} = 1$. 
The notation "sinn" means the standard function
$\sinh$ for $\Omega_k > 0$, and $\sin$ for $\Omega_k < 0$, respectively. If $\Omega_k = 0$, then "sinn $x$" means $x$ and
one simply has 
\begin{equation}
\frac{H_{0} d_{PM}(z)}{c} =  \int_{0}^{z} \frac{dz'}{\sqrt{(1+z')^3 \Omega_M  + \Omega_{\Lambda}}} \;,
\end{equation}
where $\Omega_M + \Omega_{\Lambda} = 1$.
The proper-motion $d_{PM}(z)$ distance is connected to luminosity distance $d_L(z)$ by relation
$d_{PM}(z) (1+z) = d_L(z)$ \citep{wei72}.  In both equations, it must be $\Omega_M \geq 0$. (The case
$\Omega_M = 0$ is unphysical, but it can serve as a limit.) The remaining two
omega factors can have both signs.

Eq(1) of \citet{meri13} and Eq(1) here are equivalent. Eq(3) of \citet{meri13} and Eq(2) here are
identical. 
 
In the special case of $\Omega_{\Lambda} = 0$, the integral in Eq(1) can be evaluated by 
the so-called Mattig formula \citep{mat58}. In the special case of
$\Omega_k = 0$, the elliptic integrals can be used \citep{meri13}. The special case, where
$\Omega_M = 0$, in Eq(1) the integral is very simple.  
As a result, if one omega factor is zero, but the remaining two ones are non-zeros, 
no numerical integration is needed. In addition, if just two
omega factors are simultaneously zeros, then the integration via primitive functions is very easy. 
Only the case where all three omega factors are non-zeros needs numerical integration.  
We also search for eventual analytical integrations here. 

\section{Integration}

In what follows we assume that $\Omega_{\Lambda} \neq 0$, $\Omega_k \neq 0$, $\Omega_M > 0$, and
$\Omega_k + \Omega_M + \Omega_{\Lambda} = 1$ in Eq(1).
We rewrite the denominator of Eq(1) as follows:
$$
\sqrt{(1+z')^3 \Omega_M  + (1+z')^2\Omega_k +\Omega_{\Lambda}} \;\; =
$$
\begin{equation}
\sqrt{(1+z')^3 \Omega_M  +\Omega_{\Lambda}}\times
\sqrt{1 + \frac{(1+z')^2\Omega_k}{(1+z')^3 \Omega_M  +\Omega_{\Lambda}}} \;.
\end{equation}
 
Using only the first two terms in the Taylor series for function $g(Q)$,
\begin{equation}
g(Q) = \frac{1}{\sqrt{1+Q}} = 1 - \frac{Q}{2} +  ..., 
\end{equation}
and defining 
\begin{equation}
Q =  \frac{(1+z')^2\Omega_k}{(1+z')^3 \Omega_M  +\Omega_{\Lambda}} \;,
\end{equation}
we obtain for the integral in Eq(1) 
$$
\int_{0}^{z} \frac{dz'}{\sqrt{(1+z')^3 \Omega_M  + (1+z')^2\Omega_k +\Omega_{\Lambda}}} =
$$
$$
\int_{0}^{z} \frac{dz'}{\sqrt{(1+z')^3 \Omega_M  +\Omega_{\Lambda}}}\; - \;\;\;\;\;\;\;
$$
\begin{equation}
\Omega_k \int_{0}^{z} \frac{(1+z')^2 dz'}{2 (\Omega_M (1+z')^3+ \Omega_{\Lambda})^{3/2}} + ...\;.
\end{equation}
It is essential to emphasize here that the first term on the right hand side of this equation is
not identical to the right hand side of Eq(2). The difference is given by the fact that in
Eq(2) it holds that $\Omega_M + \Omega_{\Lambda} = 1$, but here it holds that $\Omega_M + \Omega_{\Lambda} = 
1 - \Omega_k \neq 1$. 

Inspecting Eqs(3-13) of \citet{meri13}, we recognize that the constraint $\Omega_M + \Omega_{\Lambda} = 1$
is not needed from the mathematical point of view. We can easily repeat the whole 
integration procedure of Eqs(3-13) in \citet{meri13} for any $\Omega_M > 0$ and $\Omega_{\Lambda} > 0$. 
All this means that the first term on the right hand side of Eq(6) can again be solved analytically using
the elliptic integral of first kind. The only difference is given in Eq(5) of \citet{meri13}.
In \citet{meri13} $\Omega_M + \Omega_{\Lambda} = 1$ was used, but here both omega factors can be arbitrarily positive.
This means that only the factor $\Omega_M^{1/3}\Omega_{\Lambda}^{1/6}$ is influenced, but the integral in
Eq(5) of \citet{meri13} itself remains unchanged. For the sake of completeness, it must be added that, using the remark
of Section 3 of \citet{meri13}, we can generalize here the solution also for arbitrary 
$\Omega_{\Lambda} < 0$ and $\Omega_M > 0$. Hence, the first term of the right hand side of Eq(6) can always 
be calculated analytically.

On the other hand, the second term of Eq(6) is a complicated formula, and we did not find any primitive
function for this integral. It should therefore be calculated numerically. In either case, this integral is finite for any 
$0 < z < \infty$.

\section{Remarks}

It may seem that Eq(6) did not give any new result, because - in essence - some numerical integration is again necessary.
But this is not the case, because
for small $0 < |\Omega_k| \ll 1 \simeq \Omega_M + \Omega_{\Lambda}$, the procedure of the previous section can be useful.
This can be seen as follows.

The interval of convergence of the Taylor series for $g(Q)$ in Eq(4) is given by $|Q| <1$. This means that it has to be
\begin{equation}
0 < |(1+z')^3 \Omega_M  + \Omega_{\Lambda}| - (1+z')^2 |\Omega_k| = f(z')
\end{equation}
for any $\infty > z' \geq 0$. If this requirement holds, then even higher terms of the Taylor series can be used. 

The requirement of Eq(7) can also be formulated in an other form that does not use $z'$. 
For $- \Omega_{\Lambda} \geq \Omega_M > 0$, the first term in $f(z')$
is zero for $1 + z' = (-\Omega_{\Lambda}/\Omega_M)^{1/3} \geq 1$. Hence, for this $z'$ one has $f(z') < 0$, and 
$|Q| <1$ is not fulfilled. Thus, the term $(1+z')^3 \Omega_M  + \Omega_{\Lambda}$ must be positive. 
For $z' = 0$ we must also have $|Q| <1$, and it has to be $\Omega_M  + \Omega_{\Lambda} > |\Omega_k|$.
This combination of the omega factors is thus the necessary condition for $|Q| <1$. 
To obtain the sufficient condition, one has to discuss $f(z')$ as a standard cubic function.
If $2|\Omega_k|/(3 \Omega_M) < 1$ holds, then the necessary condition  
is also a sufficient one, because then $f(z')$ is an increasing function for any $z'\geq 0$.
If $2|\Omega_k|/(3 \Omega_M) \geq 1$ holds, then one must have $f(z') > 0$ for 
$1+ z' = 2|\Omega_k|/(3 \Omega_M)$, because $f(z')$ has a minimum for this $z'$.
It is positive for $\Omega_{\Lambda} > 4|\Omega_k|^3/(27 \Omega_M^2)$ and $|Q| <1$ holds;
but it is non-positive for $\Omega_{\Lambda} \leq 4|\Omega_k|^3/(27 \Omega_M^2)$ and $|Q| <1$ does not hold.
  
All this means that mainly for $0 < |\Omega_k| \ll \Omega_M + \Omega_{\Lambda} \simeq 1$ the procedure of this note is 
quite usable. (For the sake of precision it must be added that either the condition $2|\Omega_k|/(3 \Omega_M) < 1$ or the
condition $\Omega_{\Lambda} > 4|\Omega_k|^3/(27 \Omega_M^2)$ must also be fulfilled.) In this case,
the second term in Eq(6) is a small correction to the first one. Then it can be simpler to numerically calculate 
only this small correction and not the whole integral of Eq(1). In any case, our "semi-analytical" procedure 
can serve as a check of the numerical calculation of Eq(1).

\section{Conclusion}

We have shown that the integral on the right hand side of Eq(1) can be partly calculated    
analytically using the elliptic integral of the first kind even in the case, when all the three omega factors are
non-zeros. This calculation can be useful for $0 < |\Omega_k| \ll 1$.

\begin{acknowledgements}
We wish to thank A. Cappi, D. Eisenstein, B. Fulford, M. K\v{r}\'{\i}\v{z}ek, and N. Ryan for useful discussions.
This study was supported by the Grant Agency of the Czech Republic Grant
P209/10/0734, and by Creative Research Initiatives (RCMST) of MEST/NRF.
\end{acknowledgements}

\bibliographystyle{aa} 
\bibliography{references-elliptic2}

\begin{thebibliography}{4}
\expandafter\ifx\csname natexlab\endcsname\relax\def\natexlab#1{#1}\fi

\bibitem[{{Carroll} {et~al.}(1992){Carroll}, {Press}, \& {Turner}}]{car92}
{Carroll}, S.~M., {Press}, W.~H., \& {Turner}, E.~L. 1992, \araa, 30, 499

\bibitem[{{Mattig}(1958)}]{mat58}
{Mattig}, W. 1958, Astronomische Nachrichten, 284, 109

\bibitem[{{M\'esz\'aros} \& {\v{R}\'{\i}pa}(2013)}]{meri13}
{M\'esz\'aros}, A. \& {\v{R}\'{\i}pa}, J. 2013, \aap, 556, A13

\bibitem[{{Weinberg}(1972)}]{wei72}
{Weinberg}, S. 1972, {Gravitation and Cosmology: Principles and Applications of
  the General Theory of Relativity} (Wiley)

\end{thebibliography}

\end{document}